\documentclass{article}
\usepackage{spconf,amsmath,graphicx}
\usepackage{xcolor}
\usepackage{tcolorbox}
\tcbuselibrary{breakable}
\usepackage{booktabs}
\usepackage[colorlinks=true, allcolors=blue]{hyperref}
\usepackage{color, colortbl}
\definecolor{LightCyan}{rgb}{0,1,1}
\definecolor{lightcornflowerblue}{rgb}{0.6, 0.81, 0.93}
\definecolor{lightgreen}{rgb}{0.56, 0.93, 0.56}
\definecolor{brightgreen}{rgb}{0.4, 1, 0}
\DeclareFontEncoding{LS1}{}{}
\DeclareFontSubstitution{LS1}{stix}{m}{n}

\DeclareRobustCommand{\rightangle}{%
  \text{\usefont{LS1}{stix}{m}{n}\symbol{"D9}}%
}
\title{Exploring data augmentation in bias mitigation \\ against non-native-accented speech}

\name{Yuanyuan Zhang$^{1}$, Aaricia Herygers, Tanvina Patel$^1$, %Munir Georges$^{1,3}$, 
Zhengjun Yue$^1$, Odette Scharenborg$^1$}
\address{
  $^1$Multimedia Computing Group, Delft University of Technology, the Netherlands \\
  }
\copyrightnotice{979-8-3503-0689-7/23/\$31.00~\copyright2023 IEEE}
\begin{document}

%\ninept
%
\maketitle

%UNCOMMENT FOR CAMERA READY
%\def\thefootnote{*}\footnotetext{Equal contribution.}

\begin{abstract}
Automatic speech recognition (ASR) should serve every speaker, not only the majority ``standard'' speakers of a language. In order to build inclusive ASR, mitigating the bias against speaker groups who speak in a ``non-standard'' or ``diverse'' way is crucial. We aim to mitigate the bias against non-native-accented Flemish in a Flemish ASR system. Since this is a low-resource problem, we investigate the optimal type of data augmentation, i.e., speed/pitch perturbation, cross-lingual voice conversion-based methods, and SpecAugment, applied to both native Flemish and non-native-accented Flemish, for bias mitigation. The results showed that specific types of data augmentation applied to both native and non-native-accented speech improve non-native-accented ASR while applying data augmentation to the non-native-accented speech is more conducive to bias reduction. Combining both gave the largest bias reduction for human-machine interaction (HMI) as well as read-type speech.
%Automatic speech recognition (ASR) systems should serve every group of speakers, not only the majority ``standard" speakers of a language. In order to build inclusive ASR systems, mitigating the bias against speaker groups who speak in a ``non-standard'' or ``diverse'' way is crucial. We aim to mitigate the bias against non-native-accented Flemish in a state-of-the-art Flemish ASR system. Since this is a low-resource problem, we investigate the optimal type of data augmentation, i.e., speed perturbation, pitch perturbation, cross-lingual voice-conversion (VC)-based methods and SpecAugment, for both the source (native accented Flemish) and the target (non-native-accented Flemish) domain data, for bias mitigation. 
%Experimental results showed that specific types of data augmentation applied to both source and target domain data improves non-native-accented ASR, while applying data augmentation to the  target domain data is more conducive to bias reduction. Combining the best data augmentation method from both the source and target domain gave the largest bias reduction for read speech.

%\begin{tcolorbox}
%The abstract should contain about 100 to 150
%words, and should be identical to the abstract text submitted electronically
%along with the paper cover sheet.
%\end{tcolorbox}
\end{abstract}
\section{Introduction} \label{sec:intro}
Automatic speech recognition (ASR) systems should serve every group of speakers, not only the majority ``standard'' or ``norm'' speakers of a language, i.e., adult, typically highly-educated, first-language speakers of a standardized language variety, without a speech disability. In order to build inclusive ASR systems \cite{scharenborg2021webinar}, mitigating the bias against speaker groups who speak in a ``non-standard'' way, which we refer to as ``diverse'' speech and includes, e.g., children, older adults, speakers with non-native and regional accents, and speakers with a voicing disorder, is crucial. 

Although no one definition of bias exists, it is often instrumentalized as the (relative) difference in word error rates (WERs) between two speaker groups, e.g., \cite{feng2021_quantifying,patel2023using, liu2022model, dheram22_interspeech}.
While research into bias quantification and mitigation is only nascent, a growing body of results shows bias against speaker groups with different sociolinguistic backgrounds, %Lately, ASR research has seen a surge in investigations into ASR performance gaps on account of sociolinguistic factors, 
including gender \cite{tatman2017gender, garnerin2019gender, garnerin2021investigating}, non-native accents \cite{zhang-yixuan2022_comparing, zhang-yuanyuan2022_mitigating, wills2023automatic}, age \cite{patel2023using, pellegrini2012impact, hamalainen2015multilingual, kathania2020formant}, and regional speech variety \cite{feng2021_quantifying, sawalha2013effects, ngueajio2022hey, koenecke2020racial, bartelds2023making}. 

%A recent study sought to quantify such performance gaps, or `biases', in both Dutch state-of-the-art (SotA) hybrid \cite{feng2021_quantifying} and end-to-end \cite{zhang-yuanyuan2022_mitigating} ASR models. They found that the model exhibited biases against various speaker groups, including regional and non-native-accented speakers. In human-machine interaction (HMI) speech, for example, speakers from the south of the Netherlands were found to recognized more poorly than other those from other regions \cite{feng2021_quantifying}. In many cases, however, the model performed poorest for speakers from Flanders (the Dutch-speaking region in Belgium) with word error rates (WERs) up to 66\%, whereas WERs as low as 16\% were found for speakers from the Netherlands \cite{feng2021_quantifying}. A follow-up study quantified the regional biases within Flanders and found higher WERs for West Flemish and Limburgish speakers compared to East Flemish and Brabantian speakers \cite{herygers2023bias}.

%Some more text here
Researchers have been looking into different techniques to mitigate these biases. For instance, \cite{zhang-yixuan2022_comparing} compared different types of data augmentation in a state-of-the-art (SotA) hybrid model, specifically speed and volume perturbation \cite{chen2020data} and pitch shift \cite{bellettini2008reliable}, and used transfer learning, specifically fine-tuning \cite{finetune} and multi-task learning \cite{chen2015speech}, as methods to reduce biases against child and adult non-native speakers of Dutch. They found that speed perturbation and pitch shift are beneficial to bias reduction, while the transfer learning techniques and volume perturbation were not.
%They found that \colorbox{yellow}{\textbf{INSERT WHAT THEY FOUND HERE}} %%Maybe also add that they also used the TDNN(F)-HMM. 
Similarly, \cite{zhang-yuanyuan2022_mitigating} examined the effects of using speed-perturbed data in addition to synthetic non-native-accented speech generated using a newly developed cross-linguistic voice conversion (VC) approach \cite{chen2021_again-vc}, and compared fine-tuning and domain adversarial training (DAT) \cite{ganin2016domain} in a SotA end-to-end (E2E) model. The inclusion of augmented and synthetic speech data resulted in lower WERs for both native and non-native-accented speech, as well as a non-nativeness bias reduction. Additionally, speed pertubation and voice transformation were shown to improve the recognition of non-native-accented English \cite{fukuda2018data}. Data augmentation can thus help bias mitigation \cite{zhang-yixuan2022_comparing, zhang-yuanyuan2022_mitigating, fukuda2018data}. 

%; however this comes at the cost of the ASR performance on other group of speakers, i.e., native speakers
%The additional data was generated through voice conversion (VC) of English speech to `non-native' Dutch speech, using AGAIN-VC \cite{chen2021_again-vc}, an autoencoder-based model. 

%Arguably, data augmentation techniques can be divided into (at least) four categories depending on where they apply and what their output is: 
Different data augmentation techniques create different
types of artificial data based on their adaptation of the original speech. In this paper, we divide these different
techniques into four categories: 1) Warping features and masking part of the training speech signal to make the acoustic model more robust, e.g., SpecAugment \cite{Park2019specaugment}; 2) Increasing the quantity of the training data by creating artificial data that is, e.g., slower/faster or louder/softer than the original speech, e.g., using signal speed and volume perturbations \cite{chen2020data, ko2015audio}; 3) Adding ``more speakers'' with the same articulation patterns as the training data, e.g., by shifting the pitch, e.g., pitch perturbations \cite{bellettini2008reliable}; 4) Adding ``more speakers'' with new articulation patterns thus increasing the amount of variability in the training data, e.g., through voice conversion (VC) \cite{chen2021_again-vc,chen2022data,DBLP:conf/interspeech/PranantaH0S22}. Different data augmentation techniques have different effects on recognition performance and bias reduction and these effects are dependent on the ASR architecture \cite{zhang-yixuan2022_comparing, zhang-yuanyuan2022_mitigating}. There are many open questions regarding the role of data augmentation in improving low-resource, diverse speech recognition performance and bias reduction. %, e.g., what type of diverse speech or speech variety benefits most from what type of data augmentation, and whether data augmentation should be applied to only the diverse speech data or whether data augmentation of the standard language is also beneficial.

In this study, we further and systematically investigate the 1) effect of different types of data augmentation of 2) native Flemish and diverse, non-native-accented Flemish Dutch data separately and together, for bias reduction against non-native-accented Flemish \cite{herygers2023bias} 3) for read and human-machine interaction (HMI) speech in an E2E model, with the further aim to 4) understand the relationship between different types of data augmentation and the types of diverse speech. Bias reduction is defined as reducing the WER gap between native and non-native speakers while maintaining recognition performance for native speakers.

\section{Methodology}
\label{sec:method}
%Some text here
%This section outlines the used datasets and data augmentation methods. The corresponding usage and splitting of each dataset is also included.

This section outlines the used datasets, experimental setup, applied data augmentations, and evaluation metrics. All models were trained on Netherlandic and/or Flemish Dutch ``norm speech'' from the Spoken Dutch Corpus (Section~\ref{subsubsec:method:datasets:cgn}). We also used native and non-native-accented Flemish ``diverse'' speech from the JASMIN-CGN corpus for training and testing purposes (Section~\ref{subsubsec:method:datasets:jasmin}). Additionally, we used English speech data from the VCTK corpus \cite{vctk} (Section~\ref{subsubsec:method:datasets:vctk}) to generate new ``non-native'' accented speech. We created a baseline for Flemish. Subsequently, we ran two parallel sets of experiments, one in which different data augmentation techniques were applied to the native Flemish data and one in which these data augmentation techniques were applied to the non-native data to investigate the effect of adding more data and adding more data similar to the diverse data for which recognition performance and bias needs to be improved. %and add more native/non-native-accented Flemish data as well as diverse augmented data to investigate the effect of different types of data augmentation on non-native-accented speech performance and bias reduction against non-native-accented speech.%improve the recognition of non-native-accented speech.

\subsection{Datasets}
\label{subsec:method:datasets}
%Some text here

\subsubsection{Spoken Dutch Corpus (CGN)}
\label{subsubsec:method:datasets:cgn}
The Spoken Dutch Corpus (\textit{Corpus Gesproken Nederlands}, CGN) \cite{oostdijk-2000-spoken} contains Dutch speech data spoken by 18- to 60-year-old native speakers of Dutch from both the Netherlands (NL) and Flanders (FL). The type of speech ranges from lectures, read speech, and conversational telephone speech (CTS) to broadcast news (BN). The total amount of raw audio recordings in the training set is about 900h. After cutting the full audio recording into small chunks and removing the silent chunks, we obtain the 690.45h training set from the CGN corpus, denoted as CGN-NL-FL, which consists of 424.55h of Netherlandic Dutch data (CGN-NL) and 265.9h of Flemish (CGN-FL). To test our ASR systems on ``norm'' native Flemish speech, we used two Flemish test sets from CGN, i.e., a BN test set (0.4h) and a CTS test set (1.8h). The same pre-processing steps are applied to the two test sets.

\subsubsection{JASMIN-CGN}
\label{subsubsec:method:datasets:jasmin}
The JASMIN-CGN corpus \cite{cucchiarini-etal-2006-jasmin} contains Dutch (40.48h) and Flemish (25.07h) speech data. Compared with CGN which only has native adult speakers, JASMIN-CGN has more speaker group variations. It contains five speaker groups: (1) native children between the ages of 7 and 11 years, (2) native youngsters between 12 and 16 years, (3) non-native children between 7 and 16 years, (4) non-native adults between 18 and 60 years, and (5) native older adults over 60 years. The Flemish non-native speakers were mostly Francophones. This paper aims to quantify and mitigate the bias against non-native accents in Flemish ASR, so we only used the Flemish data of the JASMIN-CGN corpus, denoted by J-FL. 
%under between the ages of 7 and 16 as well as over 65 years. It also includes speech from non-native children between the ages of 7 and 16 and adults between the ages of 18 and 60.

%Furthermore, the non-native speakers come from 37 different countries such as Afghanistan, Andorra, Egypt and Spain, etc. 
J-FL contains two speaking styles (read speech and HMI speech) spoken by native (N) and non-native (NN) speakers. We split J-FL into a training set (J-FL-Train), a validation set (J-FL-Valid), and 4 test sets (HMI-N/NN: native/non-native HMI speech, Read-N/NN: native/non-native read speech). The J-FL-Train set consists of 2 subsets: native and non-native training sets, denoted by J-FL-N and J-FL-NN. Table~\ref{tab:data-split} provides the details of the J-FL data split in terms of duration (in hours), binary gender (male (M) and female (F)), and the number of speakers and utterances.

\begin{table}[ht]
    \centering
    \caption{Amount of speech data, binary gender distribution, and number of speakers and utterances for the J-FL training, validation, and test sets splits.}
    %the CGN training and JASMIN training, validation, and test sets separately.}
    \begin{tabular}{l|c|c|c|c|c}\toprule
     \textbf{Name} & \textbf{Dur (h)} & \textbf{F} & \textbf{M} & \textbf{Spk} & \textbf{Utterances}  \\\midrule
%     CGN-NL-FL & 690.45 & - & - & - & 1082670 \\
     J-FL-Train & 20.20 & 96 & 81 & 177 & 35959\\
        \hspace{1mm} \rightangle{} J-FL-N & 11.40 & 58 & 49 & 107 & 20573\\
        \hspace{1mm} \rightangle{} J-FL-NN & 8.80& 38 &32 & 70 & 15386 \\\midrule
     J-FL-Valid & 1.00 & 96 & 81 & 177 & 1798 \\\midrule
%     Dev-s (FL) & 0.94 & - & - & - & 851 \\
%     Dev-t (FL) & 1.67 & - & - & - & 2847 \\
     HMI-N & 0.58 & 9 & 9 & 18 & 1095 \\
     HMI-NN & 0.71 & 6 & 6 & 12 & 1240 \\
     Read-N & 1.52 & 9 & 9 & 18 & 2698 \\
     Read-NN & 1.08 & 6 & 6 & 12 & 1823 \\\bottomrule
          
    \end{tabular}
    
    \label{tab:data-split}
\end{table}
\subsubsection{VCTK Corpus}
\label{subsubsec:method:datasets:vctk}
The VCTK corpus \cite{vctk} is a non-parallel English corpus consisting of studio-quality speech from 109 native English speakers (both female and male) with various regional accents. These include accents from England, Scotland, Wales, and Northern Ireland. It contains approximately 44h of read speech. The VCTK corpus is used to train the VC models and generate non-native-accented Flemish speech data.

 \begin{figure*}[t]
    \centering
    \includegraphics[width=\textwidth]{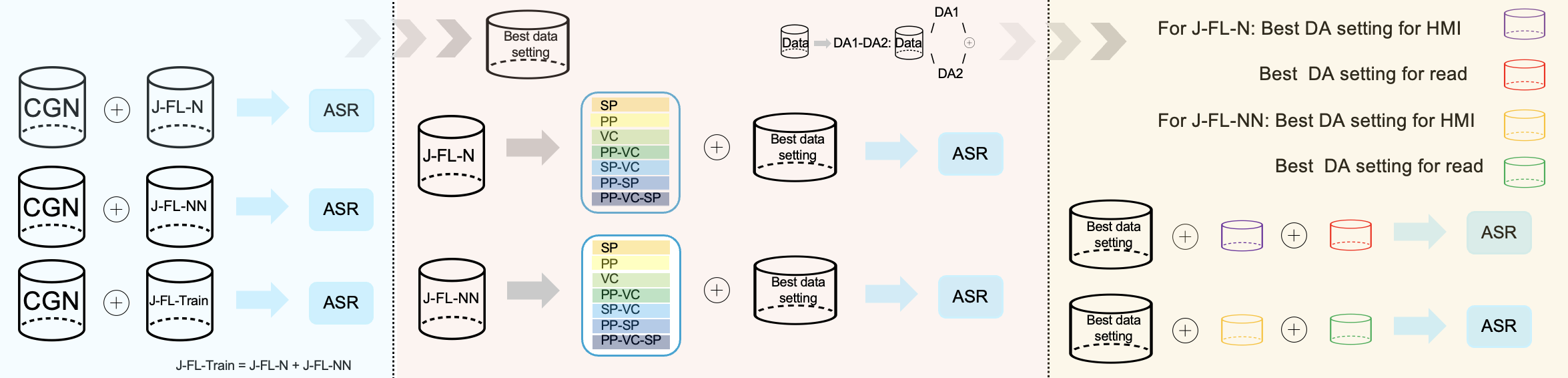}
    \caption{The experimental pipeline. The plus sign indicates the concatenation of the data sets (left panel). The best model in terms of lower bias/WERs is used for the 7 data augmentation (DA) experiments (see Section~\ref{subsubsec:method:exps:dataaug}) applied to the native and non-native data separately, for read and HMI speech separately (middle panel). The best DA methods for the native and non-native data are combined for the 2 speech styles separately in the final experiment (right panel; see Section~\ref{subsubsec:method:exps:combine}).  }
    \label{fig:exp pipline}
\end{figure*}

\subsection{Data Augmentation}
\label{subsec:method:dataaug}
\textbf{SpecAugment.}
SpecAugment \cite{Park2019specaugment} aims to bring more speech variability to the training data through time warping and time/frequency masking of the input features \cite{watanabe2018espnet}. The online augmentation settings in ESPnet \cite{watanabe2018espnet} were adopted. %It is always bring improvement to speech-related tasks \cite{Park2019specaugment, 9980314}, including ASR and \cite{9980314} speech translation \cite{bahar2019using} tasks. 
%We applied SpecAugment to the baseline and the best bias mitigation model for read speech and HMI dialogue speech.%, pursuing further non-nativeness bias mitigation.

%Some text here

\textbf{Speed Perturbation.}
Speech perturbation \cite{ko2015audio} involves the process of sampling the original unprocessed speech signal again, leading to a modified time signal with a distorted tempo. We used a two-fold speed perturbation through the \textit{sox} command with 0.9 and 1.1 perturbation factors. 
%We used the standard speech perturbation\cite{ko2015audio} data augmentation method: the speed command of \textit{sox} is used to do two-fold speed perturbation data augmentation (with $0.9$ and $1.1$ perturbation factors) for both the non-native-accented read speech data and HMI data, respectively denoted by $\mathbf{sp_{rd}}$ and $\mathbf{sp_{hmi}}$. $\mathbf{sp_{all}}$ indicates the combination of $\mathbf{sp_{rd}}$ and $\mathbf{sp_{hmi}}$.

\textbf{Pitch Perturbation.}
Pitch shift changes the voice of the speaker by shifting the original speech’s pitch by ``cents'', i.e., 1/100th of a semitone, thus increasing the speaker diversity in the training data \cite{chen2020data}. For the speech data spoken by each speaker in the pre-augment data, two random pitch values from ranges 50-200 and 200-350 were selected to generate two new speakers for each of the original speakers using \textit{sox}, resulting in two-fold pitch perturbation data augmentation.  

\textbf{Cross-lingual VC-based data augmentation.}
AGAIN-VC \cite{chen2021_again-vc} was adopted for the cross-lingual VC experiments. AGAIN-VC is a SotA autoencoder-based and non-parallel VC model that disentangles the speaker and content information in the input speech data. In voice conversion, generally speaking, target speakers provide the speaker information, while source speakers provide the content information. Since we want to create non-native-accented Flemish, we used AGAIN-VC cross-lingually with accented speech data in English from the VCTK corpus as the target speech and the native/non-native-accented speech in J-FL-N/NN as the source. %Due to: for Flemish, all the English speakers in VCTK are non-native speakers, which can provide non-native speaker information to construct new non-native-accented speech data. 
We used the VC technique to augment the non-native Flemish speech data in two ways resulting in new non-native-accented speech data in Flemish, i.e., using the native Flemish speakers in J-FL-N and the non-native Flemish speakers in J-FL-NN  as the source speech, in order to investigate the effect of the source data on bias mitigation. The demos\footnote{\href{https://non-native-accented-flemish.github.io/}{\url{https://non-native-accented-flemish.github.io/}}} of generated VC data are given. The pipeline for generating cross-lingual VC-augmented data is listed below:%In both cases, the non-native-accented speech data in VCTK served as the target speech data. %Generally, we respectively used native or non-native Flemish speakers' speech data in J-FL-N or J-FL-NN as the source to generate non-native accents-like speech data. For both cases, the pipelines of cross-lingual VC-based data augmentation are the same:
\begin{enumerate}
    \item \textbf{Train VC models}: Following the training setup in \cite{chen2021_again-vc}, the VCTK corpus (target) and J-FL-N/NN (sources) were used to train two separate VC models.
    \item \textbf{Calculate speaker embeddings}: a pre-trained ConvGRU\footnote{\href{https://github.com/RF5/simple-speaker-embedding}{\url{https://github.com/RF5/simple-speaker-embedding}}} speaker embedding model was used to obtain the speaker embeddings in the VCTK corpus, J-FL-N, and J-FL-NN.
    \item \textbf{Select source and target speakers}: The cosine speaker similarity method \cite{wan2018generalized} was used to compute the speaker similarity scores between each source speaker (from J-FL-N or J-FL-NN) and each target speaker candidate (from VCTK corpus). The two target speakers from VCTK with the two highest cosine similarity values with the source speaker were used as source speakers resulting in two-fold data augmentation.
    \item \textbf{Generate VC data}: Taking the selected source and target speaker from step 3, the VC models trained in step 1 were used to generate non-native-accented data from native/non-native Flemish speakers.
\end{enumerate}

\subsection{Experiments}
First, a solid Flemish ASR baseline was constructed (Section~\ref{subsubsec:method:asr:baselines}). Subsequently, as depicted in Figure~\ref{fig:exp pipline}, we conducted three blocks of experiments, i.e., adding native-/non-native-accented Flemish to the training data (left panel; Section~\ref{subsubsec:method:exps:addition}), carrying out data augmentation on the native- and non-native-accented Flemish separately (middle panel; Section~\ref{subsubsec:method:exps:dataaug}), and combining the best data augmentation methods of native- \& non-native-accented speech (right panel).

\subsubsection{Baselines}
\label{subsubsec:method:asr:baselines}
With Flemish being a variant of Dutch, to build a SotA Flemish ASR system, we compared ASR models trained solely on Netherlandic Dutch (CGN-NL) or Flemish (CGN-FL) and on both (CGN-NL-FL). The models were evaluated on the ``standard'' native Flemish test sets (CTS and BN) from CGN. We selected the best-performing model on CTS and BN as our baseline.

Secondly, the HMI-N/NN and Read-N/NN Flemish test sets from J-FL are rather small. To investigate their validity as our test sets, we quantified the bias against the non-native-accented speech of these small test sets and compared those with the results on the full J-FL corpus (in the remainder of our experiments, the full J-FL corpus is not used, only the separate training and test sets).% By comparing the performance (e.g., the WER and the bias) on the split test sets and the full corpus, we want to check whether the good performance is simply because the split test sets being selected are easier to recognize.

%Finally, we conducted the data augmentation experiments and compared the performance of the models trained with the different data augmentation techniques, applied to the standard and the non-native-accented speech, with that of our baseline on HMI-N/NN and Read-N/NN.

\subsubsection{Adding (Target) Native/Non-native Speech Data}
\label{subsubsec:method:exps:addition}
To investigate the potential effect of the mismatch between the training (CGN) and test (JASMIN) data, we trained three models on CGN (depicted as a cylinder with CGN in Figure~\ref{fig:exp pipline}) to which the Flemish native (J-FL-N), non-native (J-FL-NN) or both (J-FL-Train) were added. The best-performing models for HMI-N/NN and Read-N/NN were used to conduct the subsequent data augmentation experiments. Moreover, these models are compared to those in Section~\ref{subsubsec:method:exps:dataaug} to untangle the effect of the specific data augmentation technique and the effect of adding more training data. 

%As shown in Figure~\ref{fig:exp pipline}, we used the "Best data setting", i.e., the CGN data setting plus the J-FL training data setting to conduct the following experiments.

%\vspace{-1mm}
%\begin{figure}[hb]
%  \centering
%  \includegraphics[scale=0.5]{exps.png}
%  \caption{Experimental setup for native/non-%native data addition}
%  \label{fig:example transformer baseline}
%\vspace{-2mm}
%\end{figure}

\subsubsection{Data Augmentation Experiments}
\label{subsubsec:method:exps:dataaug}
We investigated the effect of the different augmentation techniques applied to the native Flemish speech J-FL-N and the non-native-accented Flemish speech J-FL-NN separately. Moreover, the data augmentation techniques were applied to both read and HMI speech. This resulted in four sets of data augmentation experiments. For the J-FL-N experiments, the best native models for read and HMI speech from Section~\ref{subsubsec:method:exps:addition} were used as the starting point. For the J-FL-NN experiments, the best non-native-accented models from Section~\ref{subsubsec:method:exps:addition} were used as the starting point. 

Speed perturbation (SP) was used to increase the variability in speech tempo in the training data which can be seen as ``simply'' increasing the amount of training data. Pitch perturbation (PP) was used to increase the ``number of speakers'' with the same accent as the original training data. The cross-lingual VC-based technique (VC) was used to add more speakers with different non-native accents compared to the original training data. The augmented data by different techniques are merged for further data augmentation (PP-VC, SP-VC, PP-SP, and PP-SP-VC). 

%different combinations of data augmentation were combined 
%For SP, PP and VC, 2-fold data augmentation was set. Moreover, the augmented data by different techniques are merged together for further data augmentation (PP-VC, SP-VC, PP-SP, and PP-SP-VC). %different combinations of data augmentation were combined 

%Moreover, based on the baseline, we added 1) FLemish native (J-FL-N), 2) Flemish non-native (J-FL-NN) and 3) both to train the ASR models. 

%1) investigate the potential effect of the mismatch in training (CGN) and test (JASMIN) sets and to 
%2) tease apart the effect of the specific data augmentation technique and the effect of simply adding more training data.

%3) both by comparing their WER and the bias against non-native-accented read and HMI speech to those of our baseline model on HMI-N/NN and Read-N/NN. 
%SpecAugment (SpecA) was used to increase the speech variability.
%The augmented data were added to train the baseline model from scratch. 
%Different models for read speech and HMI speech were trained.
%SpecAugment was applied to the best HMI and read speech bias reduction models to explore if we can get further lower biases or lower WERs.

\subsubsection{Combining Native- and Non-native-accented Data Augmentation}
\label{subsubsec:method:exps:combine}
To examine whether it is possible to further mitigate the non-nativeness bias in Flemish ASR, in the final block of experiments (see right panel in Figure~\ref{fig:exp pipline}), the best data augmentation settings for the native read speech and non-native read speech, respectively, were selected and combined resulting in a new model. The same was done for HMI speech. Subsequently, SpecAugment (SpecA) was then applied to these two new ASR models. For both read and HMI speech, the best data augmentation settings were chosen according to the lower bias, without hurting the ASR performance on native-accented speech.

%HMIdata augmentation settings that gave the lowest bias for the native and non-native-accented speech, respectively, for HMI and read speech separately.

%%need add the detailed implementation

%\colorbox{yellow}{SpecAugment \cite{Park2019specaugment}}, which is a standard augmentation method by now (needs citation)\\
%\colorbox{yellow}{Pitch Perturbation, Speed Perturbation}\\
%\colorbox{yellow}{2-fold pitch perturbation for (non-)native data}

%\colorbox{green}{VC-based non-native accents}
%VCTK \cite{vctk}
%Some text here 
\subsection{ASR Model}
\label{subsec:method:asr}
The SotA ASR model is a conformer-based sequence-to-sequence model \cite{Gulati2020conformer} in ESPnet \cite{watanabe2018espnet}. Training and testing ran on 4 NVIDIA GeForce GTX 1080 Ti GPUs. All the ASR experiments were conducted with filterbank features. 

The ASR model consists of a 12-layer conformer encoder and a transformer decoder with 5 decoder layers, all with 2048 dimensions; the attention dimension is 512 and the number of attention heads is 8. The number of batch bins is set to 10000000 for every experiment conducted in this paper. The conformer model was trained for 25 epochs using a joint connectionist temporal classification (CTC)-attention objective \cite{kim2017joint}, in which the CTC and attention weights are set to 0.3 and 0.7, respectively. 
We set the number of iterations per epoch 10000 times.
Byte Pair Encoding (BPE) units with a vocabulary size of 5000 are used as basic units. Finally, the final test model is the averaged 10-best models with 10 lowest validation losses.

\subsection{Evaluation Metrics}
\label{subsec:method:evalu}
The performance is reported in terms of WERs for the native- and non-native-accented speech and for read speech and HMI speech separately. 
%The native speech is evaluated on the Dev-S and Dev-t and the JASMIN test sets (HMI-N, Read-N). The non-native-accented speech is evaluated on the JASMIN test sets (HMI-NN, Read-NN). 
Bias against non-native accents is conceptualized as the difference between the WER performance on the native speech and the non-native-accented speech and is calculated as follows: % the difference between the WER on the non-native-accented speech and the WER on the native speech of the JASMIN corpus.
\begin{equation}
    Bias = |WER_{HMI/Read-NN} - WER_{HMI/Read-N}|
\end{equation}

%For HMI speech, the bias is denoted as Bias-HMI; for read speech, the bias is denoted as Bias-R.

\section{Results}
\label{sec:results-disc}
\subsection{Baseline Results}
\label{subsec:results-disc:baselines}
%%% move to results
%first introduce the table: say what the table shows
%2nd give a brief description of what the trends of the results are
%and then, can compare with the existing results
% two tables for two experiments, so separate them in two paragraphs
Table~\ref{tab:baselines} shows the baseline results, i.e., from training the ASR model solely on CGN-NL, CGN-FL, and CGN-NL-FL, respectively, for the conversational speech (CTS) and broadcast news (BN) CGN test sets. Bold indicates the lowest WER. Training on both the Netherlandic Dutch and Flemish (CGN-NL-FL) gave the lowest WER for both test sets, even outperforming the Flemish-only model from \cite{VanDyck2021_ASR_southerndutch}. Training the ASR only on Netherlandic Dutch gave worse results than training on only Flemish, which is not surprising given the Flemish test data. %When only training the ASR model on CGN-FL, the test results on CTS and BN are in line with previous work which demonstrates that the performance for Flemish speech is worse when training only on Netherlandic Dutch \cite{feng2021_quantifying, herygers2023bias}. 
%However, when training on CGN-NL-FL, we achieve SotA results, even better than the 
We thus move forward with our experiments using the model trained on CGN-NL-FL as our baseline. %comparing the WERs against the model trained on both NL and FL speech.

Table~\ref{tab:baseline-results-testsplit} shows the performance and bias results of the baseline model on the native and non-native speaker groups of the full JASMIN datasets and those on the JASMIN test datasets we defined for our experiments for read and HMI speech separately. The results for the Split set are slightly worse than those for the full set for read speech; however, the HMI-NN results are considerably worse for the Split set, although the results for the native speakers are slightly better for HMI speech. Overall, we conclude that the bias mitigation task is comparable in difficulty or slightly more difficult on the Split test set compared to the the full JASMIN dataset.% We can conclude that the test split from JASMIN-corpus was reasonable, which did not make the bias mitigation against non-native Flemish accents easier.
%We tested the model trained on CGN-NL-FL on the full Flemish JASMIN-CGN corpus as well as the test split outlined in Table~\ref{tab:data-split}. 

%From the results in Table~\ref{tab:baseline-results-testsplit}, the read results for the full Flemish JASMIN-CGN dataset and the split test sets are almost similar, however, the HMI-NN results (with split) are very high compared to (with full), thus there are tough non-native-accented speakers chosen to build the split HMI-NN test set and reverse is true for HMI-N. We can conclude that the test split from JASMIN-corpus is reasonable, which did not make the bias mitigation against non-native Flemish accents easier.
%does not positively influence the WERs or biases. Thus, the split is reasonable and we move further with it. 

%\begin{table}[ht]
%    \centering
%    \caption{WER and bias size of the baseline tested on the full Flemish JASMIN-CGN dataset and the split test set.}
%    \resizebox{\columnwidth}{!}{%
%    \begin{tabular}{l|c|c|c|c|c|c}\toprule
%        \textbf{Test set}  & \textbf{HMI-N} & \textbf{HMI-NN} & \textbf{Bias-HMI} & \textbf{R-N} & \textbf{R-NN} & \textbf{Bias-R} \\\midrule
%        Full & 40.0 & 52.3 & 12.3 & 27.8 & 48.4 & 20.6 \\
%        Split & 37.8 & 61.2 & 23.4 & 30.2 & 51.8 & 21.6 \\\bottomrule
%    \end{tabular}%
%    }
    
%    \label{tab:baseline-results-testsplit}
%\end{table}

\begin{table}[ht]
\vspace{-3mm}
    \centering
    \caption{WERs of the baseline systems tested on Flemish norm speech from the CGN. Bold indicates best results.}
    \begin{tabular}{l|c|c}\toprule
        \textbf{Training data}  & \textbf{CTS} & \textbf{BN} \\\midrule
        CGN-NL & 22.2 & 52.0 \\
        CGN-FL & 11.2 & 35.2 \\
        CGN-NL-FL & \textbf{8.9} & \textbf{32.8} \\\bottomrule
    \end{tabular}
    \label{tab:baselines}
\vspace{-3mm}
\end{table}

\begin{table}[ht]
\vspace{-3mm}
    \centering
    \caption{WERs and biases of the best baseline system on the full J-FL and the split test sets from the JASMIN corpus.}
    \resizebox{0.9\columnwidth}{!}{%
    \begin{tabular}{l|c|c|c|c|c|c}\toprule
       & \multicolumn{3}{c
       |}{\textbf{Read}}  & \multicolumn{3}{c}{\textbf{HMI}} \\\midrule
      \textbf{Test set}   & \textbf{N} & \textbf{NN} & \textbf{Bias} & \textbf{N} & \textbf{NN} & \textbf{Bias} \\\midrule
        Full & \textbf{27.8} & \textbf{48.4} & \textbf{20.6}& 40.0 & \textbf{52.3} & \textbf{12.3}  \\
        Split & 30.2 & 51.8 & 21.6 & \textbf{37.8} & 61.2 & 23.4 \\\bottomrule
    \end{tabular}%
    }
    
    \label{tab:baseline-results-testsplit}
\end{table}
\vspace{-3mm}

%\begin{figure*}[t]
%    \centering
%    \includegraphics[width=0.8\textwidth]{results-v6-HMI.pdf}
%    \caption{Distributions of WERs for selected models.}[Comment from Odette: Add details to boxplots: details are what the box plots contains, see other papers of Odette]
%    \label{fig:results}
%\end{figure*}
%\begin{figure*}[t]
%    \centering
%    \includegraphics[width=0.8\textwidth]{results-v6-read.pdf}
%    \caption{Distributions of WERs for selected models.}[Comment from Odette: Add details to boxplots]
 %   \label{fig:results}
%\end{figure*}

\begin{table}[ht]
    \centering
        \caption{Results of the data augmentation experiments. The shaded rows denote the lowest non-nativeness bias for read speech (blue) and HMI speech (green) in the native- and non-native-accented experiments. Best-R denotes the combination of the augmented settings of the two blue rows; Best-H denotes the combination of the augmentation settings of the two green rows.}
        %native-based experiments do not reduce the non-nativeness bias for HMI speech, so it is just (NN) 2-fold-PP.
    \resizebox{\columnwidth}{!}{% 
    \begin{tabular}{l||c|c|c||c|c|c}\toprule
       & \multicolumn{3}{c
       ||}{\textbf{Read}}  & \multicolumn{3}{c}{\textbf{HMI}} \\\midrule
      \textbf{Training Data}   & \textbf{N} & \textbf{NN} & \textbf{Bias} & \textbf{N} & \textbf{NN} & \textbf{Bias} \\\midrule
        Base & 30.2 & 51.8 & 21.6&37.8 & 61.2 & 23.4  \\
        Base w. SpecA &22.0 & 46.2 & 24.2& 33.8 & 59.1 & 26.3  \\
        \toprule
        \multicolumn{7}{c}{N/NN Data Addition}\\
        \toprule
        Base + J-FL-N & 7.2 & 25.4 & 18.2& \textbf{19.8} & 39.7 & 19.9  \\
        Base + J-FL-NN & 10.0 & 12.5 & \textbf{2.5}& 24.1 & 21.7 & 2.4  \\
        Base + J-FL-Train & \textbf{6.1} & \textbf{11} & 4.9& 20.5 & \textbf{20.7} & \textbf{0.2}  \\
        \toprule
        \multicolumn{7}{c}{Data Augmentation on J-FL-N}\\
        \toprule
        (N) SP & 4.9 & 10.0 & 5.1& 19.5 & 20.2 & 1.7  \\
        (N) PP & 5.0 & 9.8 & 4.8& 18.3 & 19.9 & 1.6  \\
        (N) VC & 4.8 & 10.1 & 5.3& 19.0 & 20.7 & 1.7  \\
        \rowcolor{lightcornflowerblue}
        (N) PP-VC & 4.2 & \textbf{8.9} & \textbf{4.7}& 18.6 & 19.7 & 0.9  \\
        (N) SP-VC & 4.2 & 9.1 & 4.9& \textbf{17.9} & 20.6 & 2.7  \\
        \rowcolor{lightgreen}
        (N) PP-SP & 4.6 & 9.5 & 4.9& 18.5 & \textbf{19.3} & \textbf{0.8}  \\
        (N) PP-VC-SP & \textbf{4.0} & 9.0 & 5.0& 18.4 & 20.5 & 1.9  \\
        \toprule
        \multicolumn{7}{c}{Data Augmentation on J-FL-NN}\\
        \toprule
        (NN) SP  & 5.7 & 9.4 & 3.7& 19.2 & 19.3 & \textbf{0.1} \\
        \rowcolor{lightgreen}
        (NN) PP& 5.7 & 9.3 & 3.6 & 19.0 & \textbf{18.9} & \textbf{0.1}  \\
        (NN) VC & 5.3 & 9.4 & 4.1& 19.4 & 20.1 & 0.7  \\
        (NN) PP-VC& 5.1 & 8.5 & 3.4 & 19.8 & 20.0 & 0.2  \\
        (NN) SP-VC& 4.9 & 8.4 & 3.5 & \textbf{18.4} & 19.2 & 0.8  \\
        (NN) PP-SP & 5.3 & 8.6 & 3.3& 18.7 & 20.0 & 1.3  \\
        \rowcolor{lightcornflowerblue}
        (NN) PP-VC-SP & \textbf{4.8} & \textbf{8.0} & \textbf{3.2}& 19.2 & 19.5 & 0.3  \\
        \toprule
                \multicolumn{7}{c}{Combining N and NN Data Augmentation }\\
        \toprule
        Best-R&3.9&7.5&3.6&19.0&18.4&\textbf{0.6}\\
        Best-R w. SpecA&\textbf{3.7}&\textbf{6.6}&\textbf{2.9}&\textbf{18.7}&\textbf{17.0}&1.7\\
        \toprule
        %(NN)PP w. SpecA& 19.0&	17.3&	1.7&	4.4	& 7.0&	\textbf{2.6}\\
        Best-H &	4.5	&8.6&	4.1&17.6	&18.9	&1.3\\
        Best-H w. SpecA &\textbf{3.4}&	\textbf{6.9}&	\textbf{3.5}& \textbf{17.5}&	\textbf{17.4}&	\textbf{0.1}	\\
        \bottomrule
    \end{tabular}%
    }
    \label{tab:data-augmentation-results-all}
\end{table}

\subsection{Data Augmentation Results}
\label{subsec:exp-results}
%1st describe the table
Table~\ref{tab:data-augmentation-results-all} shows the recognition results and bias of the experiments in five blocks: 1) the baseline (same as in Table~\ref{tab:baseline-results-testsplit}), the baseline system augmented with SpecAugment; 2) the experiments with native and non-native-accented Flemish added; 3) native data augmentation experiments; 4) non-native data augmentation experiments; and 5) the combined system.
%native/non-native data addition results (described in Section~\ref{subsubsec:method:exps:addition}), the 7 data augmentation results for native accented speech data, 7 results for non-native-accented speech data (described in Section~\ref{subsubsec:method:exps:dataaug}) and for both (described in Section~\ref{subsubsec:method:exps:combine}).

%describe the results keeping in mind the main aim of this experiment, help the reader to understand what I talk about
%2nd give a brief description of what the trends of the results are
%and then, can compare with the existing results
%the main aim of this experiment: to investigate the potential effect of the mismatch in training (CGN) and test (JASMIN)

\textbf{Applying SpecAugment.} Applying SpecAugment improved recognition performance for both speaker groups and both speaking styles, but at the cost of an increased bias.

\textbf{Adding natural native and non-native data.} First, adding native, non-native-accented, or both to the CGN data (see block 2 of Table~\ref{tab:data-augmentation-results-all}) improves recognition performance substantially (compared to Base w. SpecAug). Adding the non-native-accented data gave better recognition results than adding only native data and also substantially reduced bias against non-native-accented speech. Second, the best results for HMI-N, R-N, and R-NN were obtained when both native and non-native Flemish data from the JASMIN-CGN corpus, i.e., J-FL-Train, were added to the training data.  %The comparison against the baseline shows that adding native/non-native/both data is/are effective for the improvement of the overall performance as well as the mitigation of biases against non-native accents. 
%However, the addition of non-native speech substantially decreases the bias compared to the addition of native speech. 

%The large observed bias reduction for HMI speech differs from earlier results found for Netherlandic Dutch bias mitigation \cite{zhang-yuanyuan2022_mitigating}, where adding Dutch data from the JASMIN-CGN corpus was not able to nearly remove the WER gap between native and non-native speakers, while for Flemish HMI speech, adding Flemish data from the JASMIN-Corpus to the training data did. The following experiments start from the Base+J-FL-Train model. 

%2nd give a brief description of what the trends of the results are
%and then, can compare with the existing results
%the main aim of this experiment: to tease apart the effect of the specific data augmentation technique and the effect of simply adding more training data

\textbf{Native data augmentation.}
Comparing the results of the seven data augmentation techniques against the Base + J-FL-Train results (block 3) showed that adding different types of augmented native-accented speech data individually improved recognition performance only marginally for read and HMI speech. However, for HMI speech, adding SP, PP, or VC increased the HMI bias by 1.5\%. The best results, i.e., lowest WER for bias against non-native-accented speech, for read speech were obtained when combining PP-VC (blue row) and for HMI speech when combining PP-SP (green row).
%Adding bothPP and VC applied on the native data to the training data reduced the non-nativeness bias for read speech. Besides, the addition of other combinations of augmented native accented speech did not reduce any non-nativeness bias because they improved HMI/Read-N more than HMI/Read-NN.

%2nd give a brief description of what the trends of the results are
%and then, can compare with the existing results
%the main aim of this experiment: to tease apart the effect of the specific data augmentation technique and the effect of simply adding more training data

\textbf{Non-native data augmentation.}
Comparing the results of the seven data augmentation techniques against the Base + J-FL-Train results showed that adding different types of augmented non-native-accented speech data individually improved recognition performance again marginally for both read and HMI speech. PP augmentation yielded the lowest WER for the non-native-accented speech for HMI speech and the lowest bias. Combining all data augmentation techniques gave the best recognition and bias results for read speech. %SP/PP augmented J-FL-NN data reduced the non-nativeness bias and WERs for both read and HMI speech. With PP, for HMI speech, the recognition performance on non-native-accented speech was even better than that on native speech, resulting in only 0.1\% bias (the lowest bias for HMI in this paper). Adding more speakers with new non-native accents reduced the non-nativeness bias for read speech but enlarged that for HMI speech. PP-VC, SP-VC, and PP-SP combinations further reduced the bias for read speech a bit but not the case for HMI speech. The addition of all the three types non-native data augmentation gave the lowest WERs and bias for read speech. 

\textbf{Combining the best native and non-native data augmentation approaches.}
%1st decribe best-R w/o. SpecA, 
%and then Best_H w/o. SpecA, , 
%compring with the shaded results
%compring with baseline w. SpecA
Combining the best augmentation approach for the native- and non-native-accented speech for read speech (blue rows; Best-R) further reduced the recognition performance and bias for read speech and for the non-native-accented HMI speech and bias. Adding SpecAugment gave the overall best results with the lowest recognition rates for both N and NN speech, and with the smallest bias for read speech at the cost of a slight increase in the bias for HMI. Combining the best augmentation approaches for HMI speech (green rows; Best-H) also improved recognition performance for both speaker groups and speech types. Adding SpecAugment further improved both recognition performance and bias. Interestingly, the models that combined the best augmentation strategies for read speech and HMI speech, respectively, also improved recognition performance for the speech type on which it was not optimized. This suggests that adding more artificial data helps in both recognition performance and bias reduction.

A comparison across all models shows that the best results were obtained when adding data through multiple augmentation techniques. We thus show that pitch perturbations (PP) and speed perturbations (SP) of native- and non-native-accented data, and SpecAugment, where also adding native- and non-native-accented voice-converted data improved native and non-native-accented speech recognition and reduced bias against non-native-accented speech. 

\vspace{-1mm}
\begin{figure}[hb]
  \centering
  \includegraphics[scale=0.38]{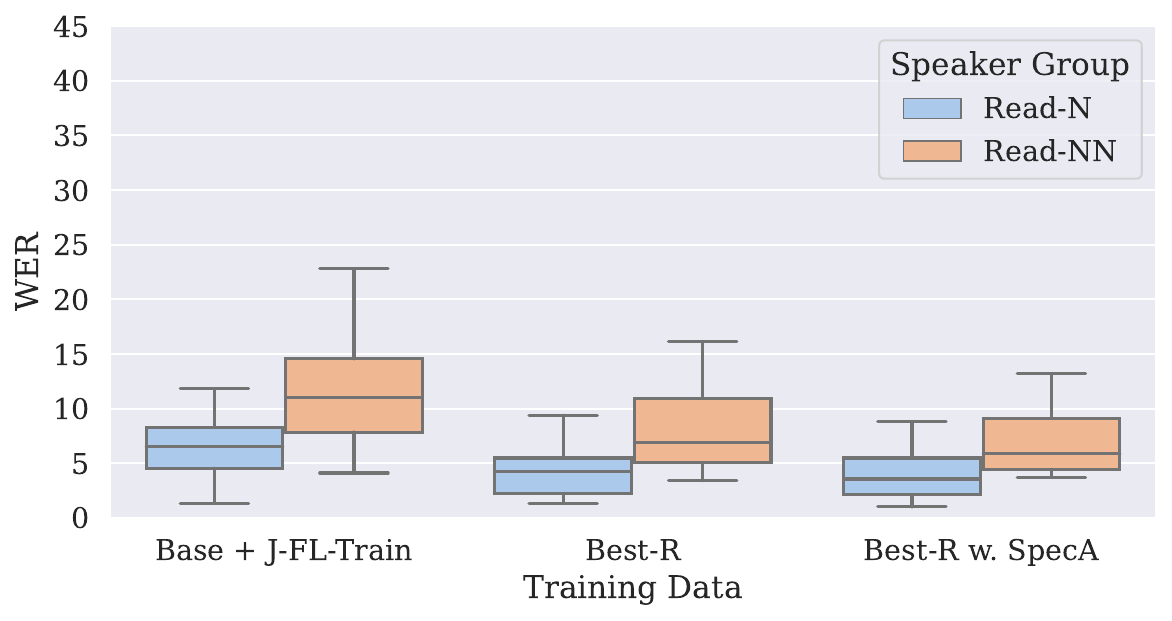}
  \caption{Distributions of WERs for Read-N/NN.}
  \label{fig:results-read}
\vspace{-2mm}
\end{figure}
\vspace{-1mm}
\begin{figure}[hb]
  \centering
  \includegraphics[scale=0.38]{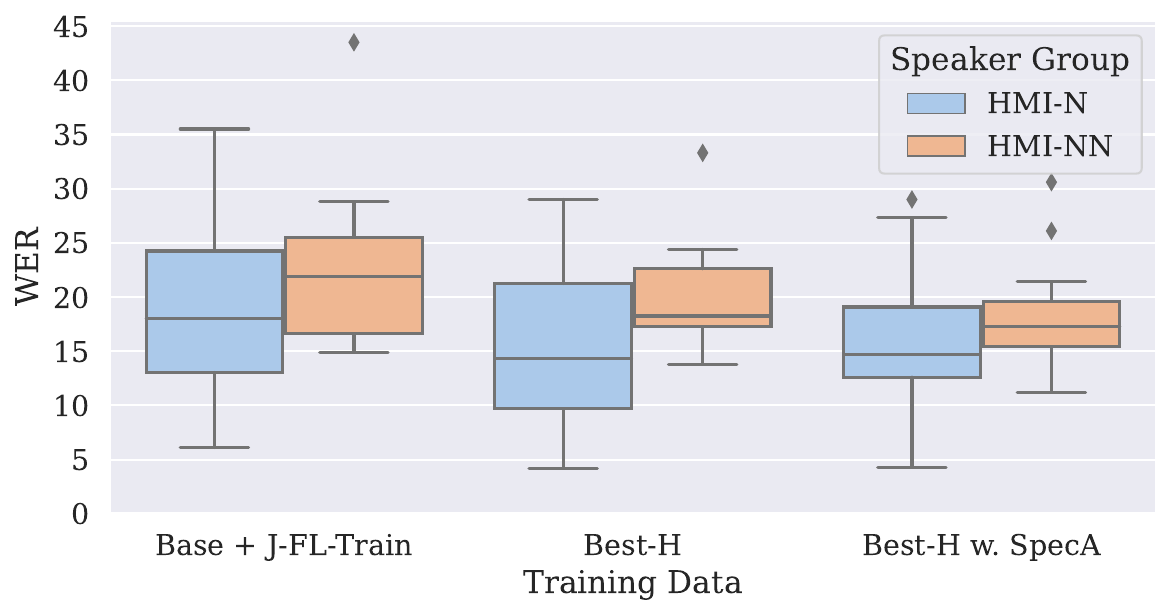}
  \caption{Distributions of WERs for HMI-N/NN.}
  \label{fig:results-hmi}
\vspace{-2mm}
\end{figure}

%Comparing the results of the combined 2-best data augmentation settings for read speech (with Best-R) and the 2 best settings separately (shaded in blue in Table~\ref{tab:data-augmentation-results-all}), there were not shown further bias reduction but improved the WERs for non-native-accented speech and read native speech. With SpecA, the lowest bias (2.9\%) ASR model for read speech was achieved. Comparing the results of the combination experiment for HMI speech (with Best-H) and the 2 best settings separately (shaded in green in Table~\ref{tab:data-augmentation-results-all}), the bias became larger because the WER on native HMI speech reduced from 18.5\% to 17.6\%. With SpecA, compared with Best-H, the WER on non-native HMI speech was improved by 1.5\% maintaining the WER on native HMI speech, resulting in 0.1\% bias. Comparing applying SpecA on baseline (it reduced WERs on the four test sets but enlarged bias), applying SpecA on Best-R and Best-H leaded to bias reduction for read and HMI speech respectively.

Figures~\ref{fig:results-read} and \ref{fig:results-hmi} show the distributions of the WERs for the nine native and six non-native speakers from our test sets across selected models, i.e., the model trained on (CGN-NL-NL and J-FL-Train) and the respective best models w/o. SpecA. They illustrate how the models with augmented data are not only able to reduce the overall WERs and biases, but also decrease the size of the WER distributions.% and eliminate outliers. 
The results provided in Table~\ref{tab:data-augmentation-results-all} show that there was no one data augmentation method/combination of data augmentation methods that gave the lowest WERs/bias across the board. However, the lowest bias for read and HMI speech can be achieved separately in two ASR models. Furthermore, we showed that more data always resulted in better performance.

\section{Discussion and Conclusion}
\label{sec:disc-conc}
In this study we systematically investigated the effect of different types of data augmentation and the type of data it was applied to for non-nativeness bias reduction. The experimental results indicate that, overall, adding more data, whether natural (see Block 2 of Table 4) or artificial (Blocks 3-5), helps reduce WERs for native- and non-native-accented speech and the non-nativeness bias. Interestingly, adding artificial non-native data created using native speakers also improves non-native-accented speech recognition (Block 3, VC), but at the cost of a small increase in bias when compared to a model trained on a small amount of natural non-native-accented speech (Block 2). If a small amount of non-native-accented speech data is available, it is preferable to use that for data augmentation as adding artificial non-native-accented data slightly outperforms adding native artificial data. The type of data augmentation does not really seem to matter, as all techniques reduced the bias and improved the performance, though different combinations do yield slightly different results. %However, for gender bias, adding more female data reduced the bias but did not improve the overall WER \cite{garnerin2021investigating}.%what are the overall trends here?
This paper extends the growing body of work investigating non-native speech recognition, e.g., \cite{fukuda2018data, wills2023automatic}. Specifically, we expand on the findings for bias mitigation against non-native-accented Dutch \cite{zhang-yixuan2022_comparing, zhang-yuanyuan2022_mitigating}. We not only showed that SP, PP, and VC helped reduce bias against non-native-accented Flemish, found in \cite{herygers2023bias}, but also showed that PP outperforms SP \cite{zhang-yixuan2022_comparing}. Comparably to \cite{fukuda2018data}, we found that SP outperforms VC \cite{zhang-yuanyuan2022_mitigating}. 
%Here, we extend the findings for bias mitigation against non-native-accented Dutch \cite{zhang-yixuan2022_comparing, zhang-yuanyuan2022_mitigating} and non-native-accented US English speech recognition \cite{fukuda2018data}. We not only showed that SP, PP and VC helped reduce bias against non-native-accented Flemish, but also that PP outperforms SP \cite{zhang-yixuan2022_comparing}, and SP outperforms VC \cite{zhang-yuanyuan2022_mitigating} and voice transformation \cite{fukuda2018data}. 
This suggests that adding ``more speakers'' might be more beneficial than adding more data. The lower results for VC might be due to the quality of the generated speech. Future work may look into the effect of VC quality on bias reduction. 
%\cite{wills2023automatic} also investigated non-native-accented (child) Dutch using JASMIN-CGN. Though our WER results are slightly better than theirs, the results are not directly comparable due to different test sets. 

In conclusion, we found that the addition of augmented data can considerably reduce WERs and biases. With SpecAugment, adding more speed-perturbed native speech data and more speakers with the same accents as both native and non-native speakers to the training data nearly removed the bias against non-native accents for HMI speech in a SotA Flemish ASR system; adding more speakers with native and (new) non-native accents and increasing the amount of non-native-accented speech reduced the bias against non-native-accented speech for read speech from 21.6\% to 2.9\%. Applying non-native-accented speech data augmentation always led to bias reduction while combining it with native speech data augmentation further improved recognition performance without enlarging the bias. We advocate for more research on bias and specifically bias mitigation in ASR.
%In conclusion, this paper sought to systematically examine the role of various data augmentation techniques in bias mitigation for non-native-accented Flemish speech. We found that the addition of augmented data can considerably reduce WERs and biases, although there are differences in the helpfulness depending on whether the speech is read or semi-spontaneous. Future work may study the generalizability of these techniques across languages.

% \section{ACKNOWLEDGMENTS}
% \label{sec:ack}

% Do not include acknowledgments in the initial version of the paper submitted for blind review.
% If a paper is accepted, the final camera-ready version can (and probably should) include acknowledgments. 

\newpage
\bibliographystyle{IEEEbib}
\bibliography{refs}

\end{document}